\documentclass{astlb}

\usepackage{graphicx}
\usepackage{natbib}
\usepackage{color}

\usepackage[hyperfootnotes=false]{hyperref}

\usepackage{apjfonts}

\usepackage{multirow}

\definecolor{darkblue}{rgb}{0,0,0.9}

\begin{document}

\journalinfo{2012}{38}{4}{238}[248]

\title{\large On the Change of the Inner Boundary of an Optically Thick Accretion
Disk around White Dwarfs Using the Dwarf Nova SS Cyg as an Example}

\author{M. G. Revnivtsev\email{revnivtsev@iki.rssi.ru}\address{1,2}, R.A.Burenin\address{1}, A. Yu. Tkachenko\address{1}, I.M.Khamitov\address{3},
T. Ak\address{3,4}, A.Merloni\address{2}, M. N. Pavlinsky\address{1}, and R. A. Sunyaev\address{1,5}
  \addresstext{1}{1Space Research Institute, Russian Academy of Sciences, Profsoyuznaya ul. 84/32, Moscow, 117997 Russia}
  \addresstext{2}{Excellence Cluster Universe, Technische Universitaet Muenchen, Boltzmann str. 2, 85748 Garching, Germany} 
  \addresstext{3}{T\"UB\"IAK National Observatory, Antalya, 07058 Turkey}
  \addresstext{4}{Istanbul University, Department of Astronomy and Space Sciences, Istanbul, Turkey}
  \addresstext{5}{5Max-Planck-Institut f\"ur Astrophysik, Karl-Schwarzschild-Str. 1, Postfach 1317, D-85741 Garching, Germany}
}

\shortauthor{REVNIVTSEV ET AL.}

\shorttitle{ON THE CHANGE OF THE INNER BOUNDARY OF ACCRETION DISK}

\submitted{Oct.7, 2011}

\begin{abstract}
We present the results of our studies of the aperiodic optical flux variability for SS Cyg, an
accreting binary systemwith a white dwarf. The main set of observational data presented here was obtained
with the ANDOR/iXon DU-888 photometer mounted on the RTT-150 telescope, which allowed a record
(for CCD photometers) time resolution up to 8 ms to be achieved. The power spectra of the source's flux
variability have revealed that the aperiodic variability contains information about the inner boundary of the
optically thick flow in the binary system. We show that the inner boundary of the optically thick accretion
disk comes close to the white dwarf surface at the maximum of the source's bolometric light curve, i.e., at
the peak of the instantaneous accretion rate onto the white dwarf, while the optically thick accretion disk
is truncated at distances $8.5\times10^9$ cm $\sim10 R_{WD}$ in the low state. We suggest that the location of the inner
boundary of the accretion disk in the binary can be traced by studying the parameters of the power spectra
for accreting white dwarfs. In particular, this allows the mass of the accreting object to be estimated.
\end{abstract}

\section{INTRODUCTION}

Compact objects in binary systems (white dwarfs,
neutron stars, black holes) form accretion disks
around themselves from the matter coming into the
region of dominance of their gravity (see, e.g., Frank
et al. 2002). Moving toward the compact object,
the matter in the accretion disk releases gravitational
energy and heats up to high temperatures, radiating
in the optical, ultraviolet, and X-ray bands (see, e.g.,
\citealt{ss73}). It has long been pointed
out that the accretion flow around compact objects
can be inhomogeneous and consist of several regions
with significantly differing physical properties, for
example, optically thick and optically thin/corona regions
\citep{shapiro76,ichimaru77,poutanen96,done07}.

The so-called dwarf novae, accreting white dwarfs
in binary systems with low-mass Roche-lobe-filling
companion stars, are among the best-studied systems
with accretion disks. An ever-increasing set of
measurements exists for them, suggesting that the
accretion disk in some states of the binary system
is not optically thick over its entire extent \citep{pringle86,livio92,king97,gaensicke99,belle03,linnel05}. In such states, the optically thick
accretion disk ends at a certain distance from the
white dwarf within which the flow becomes optically
thin. Both the interaction with the white dwarf magnetosphere,
if the white dwarf has a sufficiently strong
magnetic field \citep{patterson94,belle03}, and
gradual evaporation of the accretion disk \citep{meyer94} can be responsible for
this transition.

However, the current status of the accretion disk
models does not allow the radius at which the optically
thick flow evaporates to be predicted with confidence
(see, e.g., \citealt{meyer01}).
Thus, the question about the truncation (evaporation)
of the inner parts of the accretion disks has
not been ultimately resolved. An additional reliable
measurement of the inner radii for the accretion disks
of accreting systems in various states would be an
important confirmation of the validity of the existing
model for the accretion flow in binary systems.

Recently, it has been shown that the inner boundary
of the optically thick disk can be estimated by
analysing the power spectrum of aperiodic noise in
the accreting source \citep{revnivtsev09,revnivtsev10}.
In this paper, we applied this method to determine the
radius at which the accretion disk in SS Cyg, a binary
system with an accreting white dwarf, is truncated.

\section{APERIODIC NOISE OF ACCRETING
SOURCES FOR DIAGNOSING THE GEOMETRY OF ACCRETION FLOWS}

It has been known almost since the discovery

of accreting binary systems that, as a rule, they
exhibit aperiodic flux variability in a wide range

of time scales--flickering (see, e.g., \citealt{linnell50,bruch92}). 

As recent studies have shown, a modulation
of the instantaneous accretion rate at different
distances from the central object and their subsequent
multiplicative addition (the so-called model of
propagating fluctuations 
\citealt{lyubarskii97,churazov01})
 are the most probable mechanism of
flickering in the light curves of accreting systems.
In this model, a variable emission is generated in
the central regions of the accretion flow (which is
confirmed, for example, by the curves of the variability
amplitude in eclipsing systems (see \citealt{bruch00,baptista04}) for accreting white
dwarfs and is shown by the fast (up to time scales
of the order of tens of milliseconds) variability of the
large fraction of their X-ray flux for accreting neutron
stars and black holes). However, the accretion
flow itself is modulated at various (including) large
distances from the compact object as a result of
the stochastic nature of viscosity in accretion disks
(see, e.g., \citealt{balbus91,brandenburg95,hirose06}), with the variations
on shorter time scales that emerge in the inner disk
regions modulating the accretion rate of the matter
coming into this region from the outer regions.

In this model, the emerging broadband power
spectrum must have features in the range of frequencies
corresponding to the characteristic time scales at
the accretion disk edges. In particular, the truncation
of the accretion disk in its central part in the model
of propagating fluctuations must lead to a break
in the variability power spectrum: the source's flux
variability must be suppressed at higher frequencies
relative to the situation where the accretion disk is
not truncated. As the inner disk boundary moves
toward the compact object, the break frequency must,
accordingly, increase.

An observational confirmation of this prediction
was given previously \citep{revnivtsev09}. In
particular, it was shown that accreting magnetized
neutron stars (X-ray pulsars) whose radius of the
magnetosphere (which determines the inner boundary
of the optically thick accretion disk) depends on
the current accretion rate exhibit a change of the
break frequency in the variability power spectrum
during bursts of activity, i.e., in periods of a significant
increase in the accretion rate. The dependence of
the break frequency on the object's X-ray luminosity
(accretion rate) agrees well with the theory of a dipole
neutron star magnetosphere compressed by a Keplerian
accretion disk.

If a similar situation with a change in the accretion
flow geometry occurs in accreting white dwarfs, then
one might expect a similar change in the properties
of their aperiodic variability. However, the number
of white dwarfs with a significant magnetic field (i.e.,
with a field that is capable of destroying the accretion
disk at great distances from the white dwarf surface)
exhibiting bursts of the accretion rate is very small,
the bursts are rare and irregular, and, consequently,
it is very difficult to observe such systems in the low
and high states. Such periods of an increase in the
instantaneous accretion rate are much more often
observed for systems with a weak magnetic field.
As the object most suitable for testing this prediction,
we chose the binary system SS Cyg -- a well known
dwarf nova, i.e., a binary system in which
the periods of a low accretion rate onto the compact
object are replaced by outbursts, the periods of active
accretion (see, e.g., the review by \citealt{osaki96}). It
follows from the existing model of accretion disks
around white dwarfs (see, e.g., \citealt{king97,lasota01}) that, despite the absence of a significant
magnetic field on the white dwarf in the low state,
the optically thick accretion disk in this binary system
still ends at a considerable distance from the white
dwarf ($>5-10R_{WD}$), but not through the interaction
with the stellar magnetosphere but through its evaporation
and transformation into an optically thin flow
\citep{meyer94,liu97,meyer01}. In the periods of
peak bolometric luminosity, when the current accretion
rate onto the white dwarf increases considerably,
the optically thick accretion disk comes close to the
white dwarf.

Since the X-ray flux from this system is fairly low
(several mCrab, i.e., only a few photons per second
per 1000 cm), it is very difficult to obtain a high quality
power spectrum in the required frequency
range (up to 1-5 Hz) from X-ray light curves.
To measure the properties of the object's aperiodic
variability, we used its optical observations. Although
the optically emitting regions are considerably larger
in size than the X-ray-emitting ones, their sizes
nevertheless do not exceed $\sim10^{10}$ cm and, hence, the
smoothing of the variability due to the finite time it
takes for light to traverse the emitting region plays no
major role up to frequencies of the order of several Hz.
It is in this frequency range that we carried out the
studies presented here.

\section{OPTICAL OBSERVATIONS}

\subsection{Influence of the Atmosphere on Photometric
Measurements}

The flux variations in the photometric series obtained
with ground-based telescopes are determined
by a number of factors. First of all, these include the
object's Poissonian photon counting noise. However,
the influence of this noise on the power spectrum of
the source's flux variability is fairly easy to take into
account by subtracting the constant level ($P(f)\propto
const$) produced by this noise from it.
In the case of the CCD array used in our measurements
(EM CCD), additional noise of the photometric
signal is produced by the electronic amplification
cascade in the reception channel. This is
because the signal amplification itself in the CCD array
becomes noisy due to the Poissonian variations
of the amplification cascade electrons (multiplicative
noise). However, this amplification is a random variable
whose values are uncorrelated in neighbouring
exposures. Therefore, it actually leads to a slightly
changed value of the constant level in the derived
power spectra.
The most significant factor that makes it difficult
to measure the aperiodic variability parameters for
sources is the influence of turbulence in the atmosphere.
Because of this turbulence, chaotic variations
of the refractive index always exist in the atmosphere.
This, in turn, leads to changes in the shape of the
stellar image at the telescope's focus, their jitter, and
scintillations.

To give an idea of what additional variability is
produced by the atmosphere above the RTT-150 telescope,
we present the variability power spectrum for
a nonvariable star taken on September 2, 2010, in
Fig. 1. Unfortunately, the properties of the atmosphere
producing this additional (with respect to the
Poissonian one) noise is not strictly fixed during the
night, i.e., a correction for the atmosphere can be
made by analysing the variability power spectrum for
the reference star photographed at a slightly different
time than that for the source being investigated only
with a low accuracy.

Differential photometry is a popular method for
combatting the influence of these factors. In this
method, the flux from the star being investigated is
measured not directly but in comparison with one or
more nonvariable field stars. Using the ratio of the
fluxes from the investigated and (nearby) reference
stars allows the influence of the changing atmosphere
to be taken into account, at least to a first approximation
(see, e.g., the power spectra of the photometric
series obtained with the same RTT-150 telescope;
\citealt{burenin11}). However, a necessary condition
for this method to be efficient is the requirement that
the field stars be at least brighter than the investigated
star. Otherwise, the statistical uncertainty in measuring
the brightness of the investigated star will be
determined not by the star itself but by the noise of
the comparison star.

The optical brightness of SS Cyg in the low state
is about 12 magnitudes. This allows USNO B.1
1335-0436095 at a distance of 2.06 arcmin from
SS Cyg with a brightness $R\sim11$ to be used as a
reference star. However, during outbursts (i.e., in the
periods of a high accretion rate in the binary system),
the system's brightness increases to $R\sim 8.5$, which
makes the use of USNO B.1 1335--0436095 as a
reference star nonoptimal. In this case, the star
BD+42 4190 at a distance of $\sim$4.7 arcmin with a
brightness $R\sim 8.0$ should be used.
Here, we used both the first and second reference
stars for the application of differential photometry.

\subsection{RTT-150 Observations}

\begin{figure*}[htb]
\begin{center}
\includegraphics[height=\columnwidth,bb=26 170 576 717,clip]{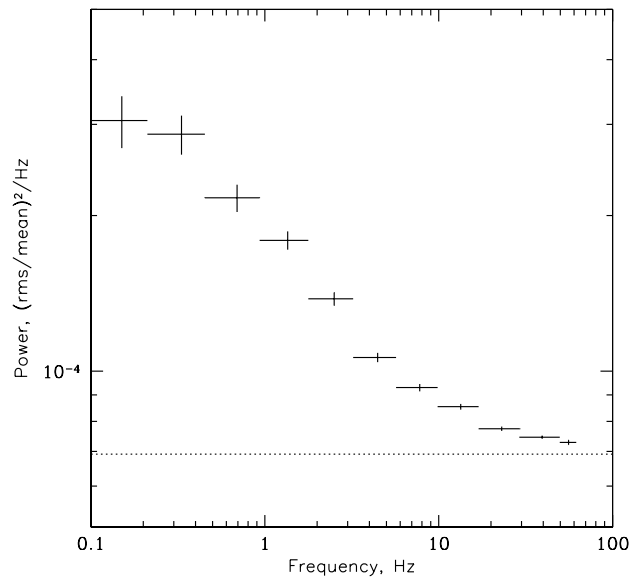}
\end{center}
\caption{Power spectrum of the flux variations for a bright nonvariable star in the series of RTT-150 observations early in
September 2010 from ANDOR/iXon measurements. We see that the variability power spectrum flattens out at frequencies
higher than several tens of Hz, approaching its ''Poissonian'' value (the estimated Poissonian noise level is indicated by the
dotted line), i.e., the noise at these frequencies is produced predominantly by the Poissonian count rate of photons (and
electrons in the amplification circuit of the ANDOR/iXon CCD array). The component in the frequency range 0.5--5 Hz
arises from atmospheric scintillations and jitter.}
\label{power_gst}
\end{figure*}

The observations of SS Cyg during its outburst
were performed with the Russian-Turkish
1.5-m telescope RTT-150 using a CCD photometer
mounted at the Cassegrain focus f = 1/7.7. A log of
observations is presented in Table 1.
The photometric measurements were made with
the ANDOR iXon DU-888 CCD array. The iXon
DU-888 back-illuminated CCD approximately 4x4 arcmin in size is divided into 1024x1024 pixels.
The CCD is equipped with electronic multiplication
(EMCCD), which allows one to reduce considerably
the readout noise effect at very short exposure times
and, consequently, to use it for measuring the brightness
of faint objects. TheCCDis cooled electronically
to a temperature of -60C.
The entire field with the readout of all CCD pixels
(1024x1024) can be imaged eight times per second;
when reducing the readout region and binning the
readout rows, the exposure time can be reduced to
$\sim$1-3 ms.

\begin{table}
\label{log}
\caption{Observations of SS Cyg on telescope RTT150 in 2010-2011, used in our work}
\medskip
\begin{center}
\begin{tabular}{l|c|c|c|c}
Date&MJD(start)& Filter& Exp.& dt\\
    &          &       & ksec &msec\\
\hline
\multicolumn{5}{c}{"Quescent" state}\\

Aug. 24, 2010&55432.8.117&g&12.3&11.7\\
\multicolumn{5}{c}{"Outburst" state}\\
Aug. 31, 2010&55439.8032&g&24.9&8.1\\
Sept.2, 2010&55441.7644&g&28.5&8.1\\
Sept.3, 2010&55442.8284&r&7.9&8.1\\
Sept.3, 2010&55443.1243&g&7.3&8.1\\
July 11, 2011&55753.8617&g&11.0&62.5\\
\end{tabular}
\end{center}
\end{table}

In our case, we used two approaches:

\begin{itemize}
\item When using USNO B.1 1335-0436095 at a
distance of 2.06 arcmin from SS Cyg as a
reference star, we recorded the image of the
sky around SS Cyg with a width of about
14 arcsec (see Fig. 2) summed in the vertical
direction into a one-dimensional strip.
This allowed us to to obtain images with a
frequency of 123.46 Hz; the length of one
exposure in our observations is 1.36 ms.
The photometric measurements were made
in a one-dimensional strip with a fixed center
and the aperture width determined from
the summed (over the interval of observations)
one-dimensional brightness profile on
the CCD. The CCD background illumination
outside bright sources was approximated by a
linear function at distances up to 100 pixels
from the sources. As the aperture width, we
took a value that was a factor of 8 larger than
the full width at half maximum of the stellar
profile. In our case, the choice of such aperture
(with a fixed value of a wide aperture) stellar
photometry is related primarily to the fact that
we needed to gather the stellar flux as completely
as possible to reduce the parasitic noise
level of the atmosphere. Reducing the aperture
width or using the aperture width determined
for each specific frame, despite the fact that
the shape of the stellar image on the detector
changes in a chaotic way unknown to us -- the
image centroid moves over the detector, the
fraction of the flux in the wings of the stellar
image changes under the atmospheric effect,
will cause the stellar flux variability amplitude
to increase through atmospheric jitter.

\begin{figure}[htb]
\begin{center}
\includegraphics[width=0.8\columnwidth]{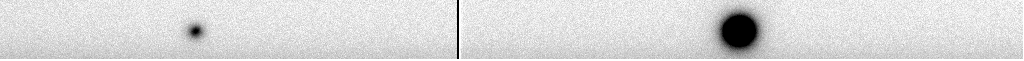}

\includegraphics[width=\columnwidth,bb=18 184 592 422,clip]{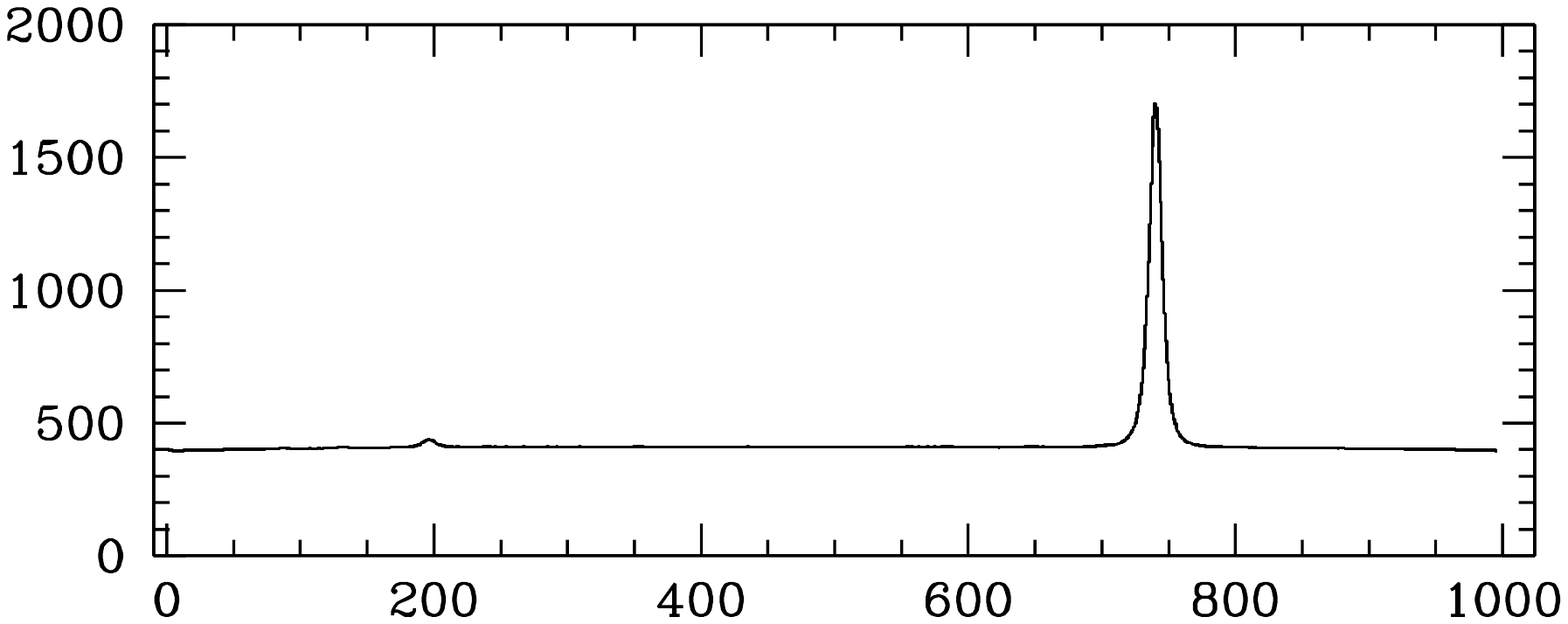}
\end{center}
\caption{Region of the sky around SS Cyg observed with the high-speed ANDOR iXon DU-888 CCD during its high state.
The comparison star (USNO B1 1335--0436095, mR = 11.0) is located in the left part of the image; SS Cyg is located in its
right part. The image with the original CCD resolution is shown at the top; the one-dimensional profile obtained by the image
addition along the vertical axis is shown at the bottom. The scale of the plot along the Y axis indicates the number of counts
recorded at the points of the one-dimensional profile in 100 s. We see that the brightness of the reference star is several times
lower than that of the investigated one.}
\label{image}
\end{figure}

\item  In 2010, we also used the method described
above during our observations of SS Cyg in
the high state, which is nonoptimal from the
standpoint of maximizing the signal-to-noise
ratio for the source due to the comparative
faintness of the reference star. To maximize
the signal-to-noise ratio in our photometric
measurements of SS Cyg in the July 2011
observation of the high state, we changed the
method and used BD+42 4190 as a reference
star. Since the CCD size is about 4 arcmin,
in this case, we cannot use the technique described
above. In this case, we placed the reference
star and the investigated source SS Cyg
along the CCD diagonal and read out the entire CCD field by binning its pixels in 2x2
(see Fig. 3). In such a case, the time resolution
was 62.5 ms. The photometric measurements
were made with a fixed (circular)
aperture. The count measurements in adjacent
circles were used as the CCD background illumination
measurements.
\end{itemize}

\begin{figure}[htb]
\begin{center}
\includegraphics[width=0.8\columnwidth,bb=129 207 507 587,clip]{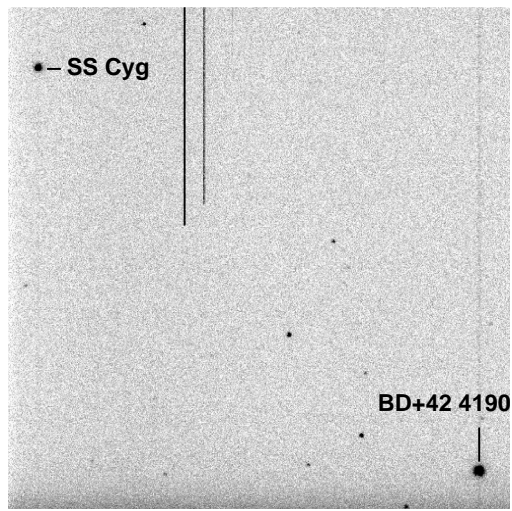}
\end{center}
\caption{Map of the sky region around SS Cyg during
its outburst in July 2011 with the reference star
BD+42 4190. The image orientation corresponds to the
CCD orientation during our observation.}
\label{image_large}
\end{figure}

\section{RESULTS}
\subsection{The Low State}

First of all, we checked whether the variability
power spectrum for SS Cyg in quiescence was consistent
with the predictions of the model of propagating
fluctuations in an accretion disk with an inner
boundary, i.e., whether there was a break in the
power spectrum similar to that observed in accreting
systems with disks truncated by the magnetospheres
of compact objects \citep{revnivtsev09,revnivtsev10}.
For this purpose, we used both RTT-150 data (see
the table) and moderate-time-resolution (about 10--30 s) observations with telescopes at the Crimean
Astrophysical Observatory (CrAO) (see \citealt{voloshina09}).

\begin{figure}[htb]
\includegraphics[width=\columnwidth,bb=26 170 576 717,clip]{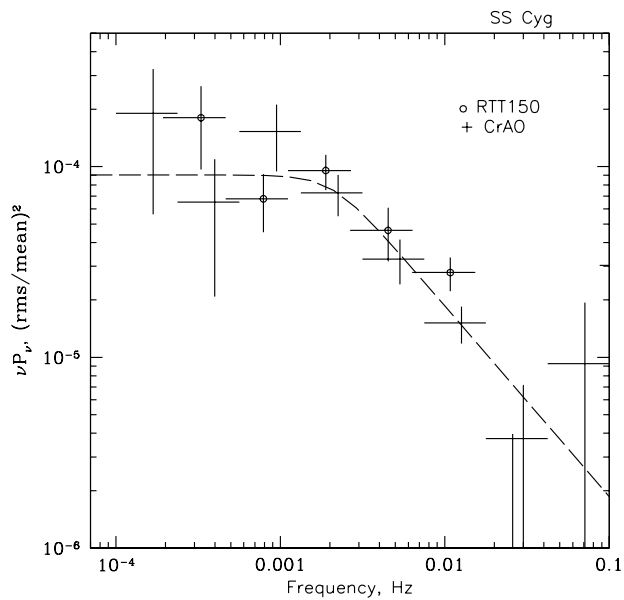}
\caption{Power spectrum of the optical flux variability for SS Cyg in the low state obtained from the RTT-150 observations in
2010 (see the text) and from the CrAO observations \citep{voloshina09}. The dashed curve indicates the model fit to the
power spectrum by a function of the form  $f\times P(f)\propto (1+(f/f_0)^4)^{-0.25}$, where $f_0=2.1\times10^{-3}$ Hz.}
\label{power_quesc}
\end{figure}

The power spectrum of the optical flux variability
for SS Cyg obtained in these observations is presented
in Fig. 4. Since the source's flux variability
is undetectable at high frequencies, Fig. 4 does not
show the frequency range above $\sim$0.1 Hz. We clearly
see that the variability power spectrum has two characteristic
regions -- the region where the variability
power behaves approximately as $P\propto f^{-1}$ (the flat
part in Fig. 4) and the region where the variability
decreases with frequency as $P\propto f^{-2}$, similar to what
we observed for intermediate polars (see Revnivtsev
et al. 2010). Note that a similar change in the pattern
of behavior of the power spectra for accreting nonmagnetic
white dwarfs was demonstrated previously
by Kraicheva et al. (1999) and Pandel et al. (2003).

\begin{figure}[htb]
\begin{center}
\includegraphics[width=\columnwidth,,bb=26 170 576 717,clip]{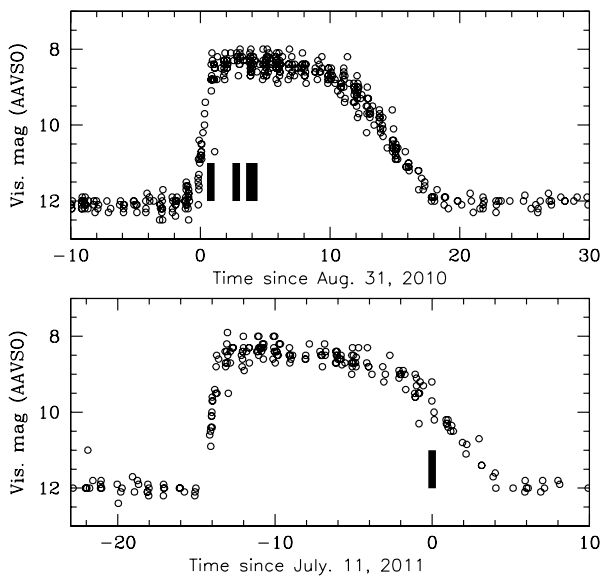}
\end{center}
\caption{Lightcurves of SS Cyg according to data of association of astronomers AAVSO (www.aavso.org) during periods Aug.-Sept. 2010 and June-July 2011. Wide black stripes denote the dates of RTT150 observations.}
\label{sep2010}
\end{figure}

If we fit the power spectrum for SS Cyg by the
analytical model $f\times P(f) \propto [1 + (f/f_0)^4]^{-0.25}$ used
in \cite{revnivtsev09}, then the break frequency
will be $(2.1 \pm  0.5) \times  10^{-3}$ Hz. For a white dwarf
with a mass of $0.81M_\odot$ \citep{bitner07}, this
corresponds to the Keplerian rotation frequency of the
matter at a distance of $(8.5 \pm 1.4) \times 10^9$ cm. The
inner disk radius estimated from the break frequency
in the power spectrum agrees satisfactorily with the
estimates based on other physical effects. In particular,
the inner radius of the optically thick disk in
the intermediate polar EX Hya estimated from the
break frequency in the power spectrum of its flux
variability, $1.9 \times 10^{9}$ cm (Revnivtsev et al. 2011),
agrees, to within about 30\%, with the results of the
measurements made by analyzing emission line profiles,
$\sim 1-2 \times 10^9$ cm (Hellier et al. 1987), and analyzing
eclipses at different phases of the white dwarf
pulsations, $\sim 1.5 \times 10^9$ cm (Siegel et al. 1989).

Interestingly, the inner boundary of the optically
thick disk determined in this way from the source?s
variability power spectrum agrees qualitatively with
the predictions of the model of evaporating accretion
disks around nonmagnetic white dwarfs (see, e.g., Liu
et al. 1997).

\subsection{The High State}

\begin{figure}[htb]
\includegraphics[width=\columnwidth,bb=26 170 576 717,clip]{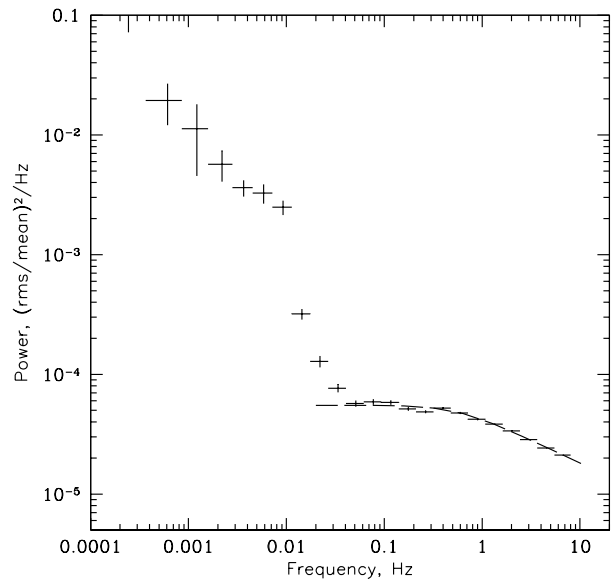}
\caption{Variability power spectrum for SS Cyg obtained by the method of differential photometry in the July 11, 2011
observation. We clearly see the residual variability of the recorded flux from the source due to an incomplete allowance for
the atmospheric jitter and scintillation effects at frequencies above $\sim 0.05$ Hz (cf. the power spectrum for a nonvariable star in Fig. 1)
}
\label{resid_atm}
\end{figure}

In the high state of SS Cyg, we made its photometric
observations in two outbursts. We managed
to obtain photometric measurements at the very peak
of the system's optical brightness in the September
2010 outburst (see Fig. 5a) and at the optical
brightness decline in the July 2011 outburst (see
Fig. 5b).

In the September 2010 outburst, we made our
measurements using USNO B.1 1335-0436095 as a
reference star. As a result, the signal-to-noise ratio of
the photometric series obtained by the method of differential
photometry is much lower than the best one
(the brightness of the reference star was lower than
that of SS Cyg by a factor of 60-70). For this reason,
the photometric series is essentially insensitive
to the possible residual atmospheric jitter effects on
the variability. Consequently, the power spectrum of
the signal from SS Cyg at frequencies above approximately
0.5-1 Hz reaches a constant that represents
the statistical measurement errors (see also the power
spectra of nonvariable stars obtained by the method
of differential photometry with RTT 150 by \citealt{burenin11}).

\begin{figure*}[htb]
\includegraphics[width=\textwidth,bb=41 491 576 717,clip]{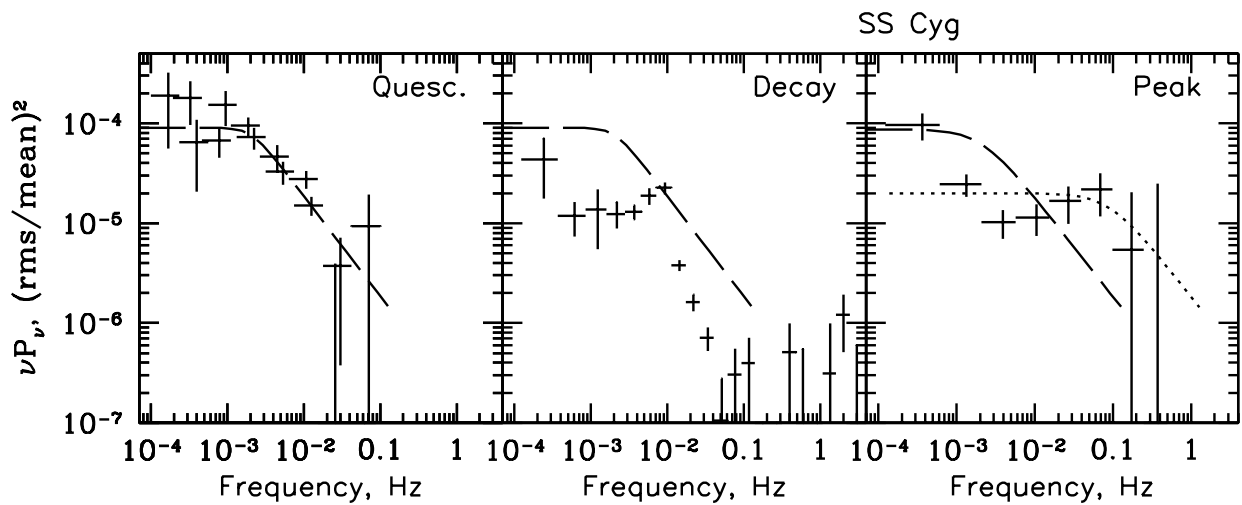}
\caption{Power spectra of the optical flux variability for SS Cyg in periods with different accretion rates in the low state (a),
in the period of an intermediate accretion rate (the July 11, 2011 observation) (b), and in the period of peak accretion rate
(the September 2, 2010 observation) (c). The dashed curve indicates the model used to describe the power spectrum in the
low state (see Fig. 4). The dotted curve indicates a similar model in which 0.091 Hz was substituted as the break frequency;
this corresponds to the Keplerian rotation frequency of matter in a circular orbit near the surface of a white dwarf with mass
$M = 0.81M_\odot$ (the white dwarf radius was calculated using a formula from Nauenberg (1972).}
\label{power_outburst}
\end{figure*}

During the July 2011 outburst, we used the
brighter star BD+42 4190. This allowed us (1) to
obtain a higher-quality variability power spectrum for
the objects and (2) to detect the residual atmospheric
jitter effects.

Figure 6 presents the power spectrum for SS Cyg
in the July 11, 2011 observation obtained by the
method of differential photometry. We clearly see the
flux variability with a shape of the power spectrum
resembling that of the atmospheric variability power
spectrum at frequencies above approximately 0.05 Hz
(see Fig. 1). Thus, we may conclude that the method
of differential photometry allows us to take into account
the atmospheric effect up to a level of about
0.3\% of the source's flux (the exact level of the residual
atmospheric jitter effect depends on the conditions in
each specific observation and on the distance between
the object and the reference star). We took into
account the influence of this residual variability by
subtracting the components in a power spectrum of
the form $P \propto A(1 + (f/f_1)^2)^\beta$. In the observation
under consideration, $A = 5.52 \times 10^{-5}$, $f1 = 0.557$,
and $\beta = -0.192$. The described model is indicated in
Fig. 6 by the dashed curve.
The power spectra of the flux variability for SS Cyg
in the light curves obtained by the method of differential
photometry during different stages of its outburst
state are presented in Fig. 7.
It should be noted that in the September 2, 2011
observations, apart from aperiodic noise (see Fig. 8),
we detected a transient appearance of the so-called
dwarf nova oscillations (DNOs) commonly observed
in SS Cyg (see, e.g., \citealt{mauche01}).
For example, in the time interval MJD 55441.7644--
55441.9581, we observed oscillations with a period
of 6.735 s and a relative amplitude of $3.2\times10^{-4}$.
However, their presence in no way affects our results
on the investigation of aperiodic variability, because
the amplitude of these oscillations is more than an
order of magnitude lower than that of the observed
aperiodic variations.
\section{CHANGES OF THE INNER DISK BOUNDARY}

The pattern of change in the aperiodic flux variability
of SS Cyg is clearly seen from Fig. 7. As
the brightness (accretion rate) of the source rose,
the fraction of the fast variability clearly increased
at frequencies up to 0.01 Hz in the July 11, 2011
observation (i.e., during the source's intermediate
brightness) and up to 0.1 Hz in the observations early
in September 2010 (i.e., during the source's peak
optical brightness).

It is in this frequency range that we should expect
the appearance of an additional noise component due
to the appearance of parts of the accretion disk near
the white dwarf surface: the rotation frequency of
matter in a Keplerian orbit near the surface of a white
dwarf with mass $M = 0.81M_\odot$ is about 0.091 Hz.
A similar appearance of the high-frequency component
in the source's flux noise was observed in
magnetic systems (in particular, accreting magnetized
neutron stars) in which the inner radius of the
accretion disk changed during the period of activity
because of the change in the balance of matter pressures
in the disk and the neutron star magnetosphere
(see Revnivtsev et al. 2009).

Thus, we may conclude that the results of our observations 
show that the accretion flow around a nonmagnetic
white dwarf (in our case, the white dwarf in
the binary system SS Cyg) is clearly divided into two
regions the location of the boundary between which
depends on the current accretion rate in the inner
part of the accretion flow (where the main energy
release takes place). In the model of an evaporating
accretion disk \citep{meyer94},
this boundary separates the regions of optically thick
and optically thin flows.

Interestingly, a distinct feature like a quasiperiodic
oscillation (QPO)with a low Q ($Q = f/\Delta f \sim
1$) is observed near the break in the variability power
spectrum for SS Cyg. The appearance of QPO
near the break in the variability power spectrum
is not unique to the object considered but is a
rather common phenomenon in systems in which
matter is accreted through two regions with different
physical properties. For example, in magnetized
neutron stars, in whichmatter fromthe optically thick
disk penetrates into the magnetosphere (see, e.g.,
\citealt{revnivtsev09}); in accreting black holes in a
hard spectral state, in which matter from the optically
thick disk transforms into an optically thin coronal
flow \citep{wijnands99}. However,
in the mentioned cases, the QPO Q factor is, as a
rule, higher. It can be assumed that the appearance
of QPO in the case of SS Cyg is associated with
instabilities whenmatter penetrates fromthe optically
thick accretion disk into the coronal flow.

The detection of a break in the variability power
spectrum for nonmagnetic white dwarfs opens up
possibilities for an independent estimation of their
masses. Indeed, the inner radius of the accretion
disk, which must exceed the white dwarf radius, can
be estimated by measuring the break frequency in
the object's variability power spectrum at the peak
of its optical-ultraviolet brightness. Next, given the
equation of state for the white dwarf (i.e., its mass--
radius relation), we can estimate its mass. Unfortunately,
in our case, the statistical quality of the
observational data at the peak accretion rate onto the
white dwarf (September 2010) is too low to independently
estimate the white dwarf mass by this method.
New, higher-quality observations are needed for this
purpose.

\section{CONCLUSIONS}

We analysed the pattern of aperiodic optical flux
variability for the accreting nonmagnetic white dwarf
SS Cyg. This work is a pilot project whose goal is to
test the hypothesis that the accretion disk around the
nonmagnetic white dwarf in SS Cyg is truncated at a
certain distance from the white dwarf in the low state
and approaches the white dwarf in the high state. For
this purpose, we carried out special observations of
the source in the low and high states with a record
(for CCD photometers) time resolution up to 123 Hz.
We showed the following.

\begin{itemize}
\item
In the low state, the power spectrum of the
optical flux variability for SS Cyg is similar
to the variability power spectrum for intermediate
polars with a break in the range of frequencies
$\sim 2\times 10^{-3}$ Hz. This suggests that
the optically thick accretion disk in SS Cyg
is truncated/evaporated (Meyer and Meyer-
Hofmeister 1994; Liu et al. 1997) at a distance
of about $(8.5 \pm 1.4) \times 10^9$ cm, or $R \sim 10R_{WD}$.

\item In the high state, the flux variability amplitude
for SS Cyg in the frequency range $10^{-3}-1$ Hz
is lower than that in the low one, 1.3\%. However,
it contains a much larger fraction of fast
variability, up to frequencies of $\sim 0.1$ Hz that
roughly correspond to the rotation frequency of
matter in a Keplerian orbit near the white dwarf
surface. In the state with the source's maximum
optical brightness, the break frequency
is maximal, about 0.1 Hz; in the state with an
intermediate brightness, the break frequency is
about 0.01 Hz.
\end{itemize}

To further study the behaviour of the inner boundary
of the optically thick accretion disk at various stages
of the burst of the accretion rate, we need to carry
out additional observations with a bright reference
star in the instrument's field of view and to try to
compare more quantitatively the inner boundary of
the accretion disk for nonmagnetic white dwarfs with
the predictions of various theoretical models.

\acknowledgements
We thank the Excellence Cluster Universe of the
Technische Universit\"at M\"unchen for the opportunity
to work with the ANDOR/iXon CCD array at the
RTT-150 telescope. M.G. Revnivtsev expresses
special gratitude to Andreas M\"uller for his great help
in acquiring the ANDOR/iXon CCD array. We used
the source's light curves measured by the American
Association of Variable Star Observers (AAVSO).
M.G. Revnivtsev is grateful to I.B. Voloshina who
provided the light curves of SS Cyg in the low
state. We thank the T\"UB\"ITAK National Observatory
(TUG, Turkey), the Space Research Institute of
the Russian Academy of Sciences, and the Kazan
State University for support in using the Russian--Turkish 1.5-m telescope (RTT-150). This work
was supported by the Russian Foundation for Basic
Research (project nos. 07-02-01004, 08-02-00974,
09-02-12384-ofi-m, 10-02-01442, 10-02-01145,
10-02-00492, 10-02-91223-ST-a), the Program for
Support of Leading Scientific Schools of the Russian
Federation (NSh-5069.2010.2), the Programs of the
Russian Academy of Sciences P-19 and OPhN-
16, The ''Dynasty'' Foundation for Noncommercial
Programs, and the T\"UB\"ITAK Programs 209T055
and 10BRTT150-25-0.

\small
\hfill {\em Translated by V.~Astakhov}

\end{document}